# Dielectric relaxation and electrical conductivity in lead-free $(100-x)(Li_{0.12}Na_{0.88})NbO_3$-$xBaTiO_3$ ($0 \leq x \leq 40$) piezoelectric ceramics: An impedance spectroscopic study

Supratim Mitra[*], Ajit R. Kulkarni

Department of Metallurgical Engineering and Materials Science, Indian Institute of Technology Bombay, Mumbai 400076, India

[*]Now at: Department of Physics, Banasthali University, Banasthali, Rajasthan 304022, India

**Abstract**

Dielectric behavior and conductivity mechanism in lead-free $(100-x)(Li_{0.12}Na_{0.88})NbO_3$-$xBaTiO_3$ ($0 \leq x \leq 40$) piezoelectric ceramics were investigated using impedance spectroscopy over a wide temperature (-100 °C to 500 °C) and frequency range (0.1 Hz to 1 MHz). The grain and grain boundary response as well as the relaxation processes at different frequencies and temperatures were also discussed. A low frequency dispersion in dielectric permittivity (LFDD), a typical characteristic of high-temperature behavior was observed both below and above the ferro-paraelectric phase transition temperature, $T_m$. Oxygen-defect-related complexes ($V^{\bullet\bullet}_O$, $(Ti_{Nb}' - V^{\bullet\bullet}_O)^{\bullet}$) generated due to acceptor type doping were found to play an important role in LFDD and hopping conduction. The dielectric relaxation follows Jonscher universal law, however LFDD is found to be associated with quasi-DC process (QCD). The activation energies of DC conduction confirm the mechanism as the thermal motion (short range hopping) of doubly ionized oxygen vacancies. At high temperature, the conductivity relaxation mechanism is dominated by grain boundary conduction through hopping electron created by the charge compensating oxygen vacancies.

# 1. Introduction

Lead-free piezoelectric ceramics are recognized as potential replacements for lead zirconium titanate (Pb, Zr)TiO$_3$ (PZT) due to the environmental concern of toxic Pb in technological industries. They have been found attractive due to good piezoelectric properties and high Curie temperature. Recently, a solid solution of lithium sodium niobate, Na$_{1-x}$Li$_x$NbO$_3$ (LNN) and BaTiO$_3$ (BT) [(100-$x$)LNN-$x$BT] has emerged as a promising lead-free piezoelectric system for resonators, filters and multilayered capacitors applications [1].

The dielectric response in these materials is mainly associated with a collective response of different microscopic polarization processes under external electric field. Under an AC field, dielectric dispersion or relaxation is observed associated with one or more polarization mechanism each characterized by different relaxation time depending on the nature of the dipoles. The dielectric relaxation in the material is greatly influenced by defects, microstructure, surface chemistry, conductivity and thus become extremely important. [2-5]. In order to understand the conduction mechanism in (100-$x$)LNN-$x$BT ceramics, the dielectric relaxation study has been carried out in frequency domain when permittivity and conductivity are concerned. This is often referred to as 'universal dielectric relaxation' that is originating from the hopping of charge carriers and with the interactions of the inherent defects in the materials. A dielectric relaxation process due to localized conduction is associated with the presence of oxygen vacancies and the nonlocalized conduction corresponding to long range conductivity is associated with extrinsic mechanisms [6].

In (100-$x$)LNN-$x$BT ceramics, A-site vacancies are most probable as A-site cations (Li$^+$, Na$^+$, Ba$^{2+}$) are highly volatile during processing and Li$^+$ may occupy the random site, which resulted in oxygen vacancies due to maintain charge neutrality. At any particular temperature, however, Gibb's free energy of a crystal is minimized when a certain fraction of ions leaves the normal lattice. As the temperature increases, more and more defects are produced which, in turn, increase the conductivity. In the high temperature (intrinsic) region, the effect of impurity on electrical conduction will not change appreciably whereas in the low temperature (extrinsic) region, the presence of impurity in the crystal increases its conductivity. The electrical conduction in dielectrics is mainly a defect controlled process in the low temperature region. The presence of impurities and vacancies mainly determines this region. The energy needed to form the defect is much larger than the energy needed for its drift. The conductivity of the crystalline material in the high temperature region is determined by the intrinsic defects caused by the thermal fluctuations in the crystal. In the low temperature ferroelectric phase (LT-FE) extrinsic conduction resulting from impurity ions present in the lattice dominates, whereas intrinsic ionic conduction results from the movement of the component ions occurs in the high temperature paraelectric phase (HT-PE) [7, 8].

Impedance spectroscopy has been proved very useful tool for investigating the relaxation behavior of polycrystalline materials arising from intragranular and interfacial regions and their interrelation. Impedance measurement helps to evaluate and separate their temperature and frequency dependent phenomena from overall electrical behavior. It indicates whether the overall resistance of the material is dominated by bulk or grain boundary and helps to measure the value of the component resistance and capacitance. It also enables us to evaluate relaxation frequency which is intrinsic to the material and independent of the sample geometry. A combined plot of impedance and modulus provide discrimination between dielectric responses that arises from localized relaxation or long range conductivity [9].

In the view of the above, the present work has undertaken a detailed study on dielectric relaxation and conductivity behavior of $BaTiO_3$ (BT) modified LNN (LNN-BT) considering universal relaxation law proposed by Jonscher. In order to obtain information about conduction mechanism associated with mobile charges, conductivity analysis was carried out.

The impedance spectroscopy technique is employed to analyze the individual contribution from grain and grain boundaries to relaxation and conduction processes.

## 2. Experimental

Solid solution of (100-$x$)Li$_{0.12}$Na$_{0.88}$NbO$_3$-$x$BaTiO$_3$ [(100-$x$)LNN-$x$BT] ($x$ = 0, 10, 20, 30, 40) was prepared using conventional solid state reaction followed by sintering. The starting raw materials for the synthesis were reagent-grade TiO$_2$, Nb$_2$O$_5$, Na$_2$CO$_3$ (both 99.5% pure, Loba Chemie, India), Ba$_2$CO$_3$ (99.0% pure, Aldrich, USA) and Li$_2$CO$_3$ (99.0% pure, Merck, India). LNN and BT powders were synthesized separately from the raw materials Nb$_2$O$_5$, Na$_2$CO$_3$, Li$_2$CO$_3$ and TiO$_2$, Ba$_2$CO$_3$ respectively (details are given elsewhere [1]. These powders were mixed in the desired stoichiometry of (1-$x$)LNN-BT (0 ≤ $x$ ≤ 0.40) and ball milled for 24h followed by compaction into pellets and pressureless sintering at different temperatures 1100-1250 °C for 3h. The phase purity of the ceramic samples was checked by powder X-Ray Diffraction (XRD) at room temperature using X-ray diffractometer (X'Pert, PANalytical) with Cu-K$\alpha$ radiation. For this, the final sintered pellets were finely crushed and annealed at 500 °C for 12h to reduce residual strain introduced by crushing. The microstructure was recorded using scanning electron microscope (SEM) (JEOL-JSM 7600F). The dielectric constant/permittivity of the unpoled samples was recorded using Impedance Analyzer (Alpha High Resolution, Novocontrol, Germany) in the frequency range 0.1 Hz-1 MHz over the temperatures -100 to 500 °C.

## 3. Results & Discussions

### 3.1. Microstructure Analysis

To study the morphology of the grains, the sintered pellets were fractured and then SEM micrographs were recorded. Figure 1 shows SEM images of fractured surfaces of (100-$x$)LNN-$x$BT ceramics for compositions, $x$ = 10, 20, 30, 40. The samples show well developed grains, clear grain boundary and overall a dense microstructure. For BT content $x$ = 10, 20, rectangular shaped grains with curved grain boundary has been observed. However, comparatively smaller and round shaped grains were observed for $x$ = 30, 40 in (100-$x$)LNN-$x$BT ceramics. The average grain size is also found to decrease with BT content ($x$) from 5.1 μm for $x$ = 0.1 to 2.1 μm for $x$ = 0.4.

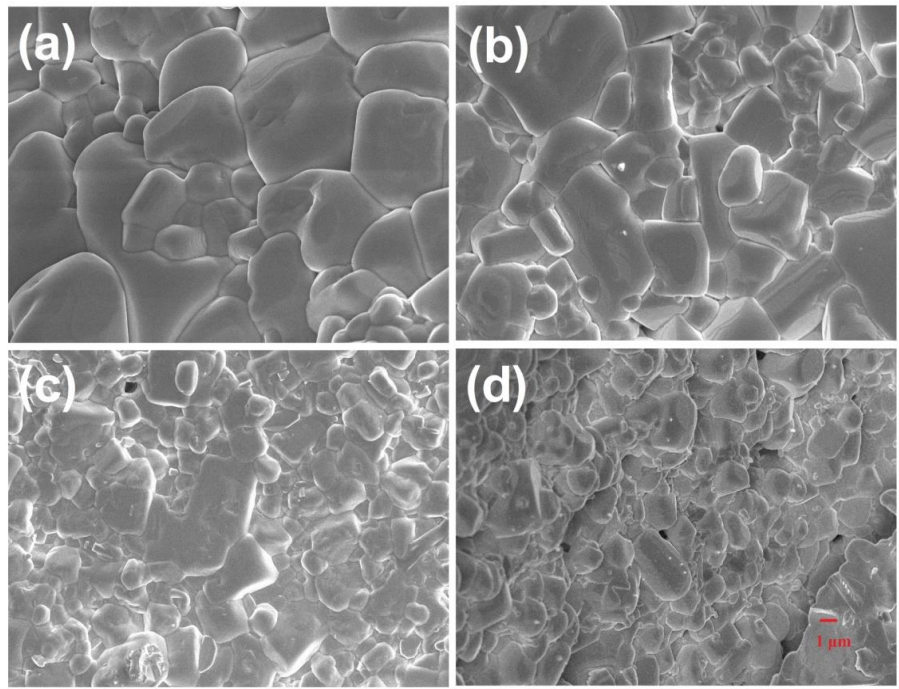

Figure 1: SEM micrographs of a fractured surface of (1-$x$)LNN-$x$BT ceramics for (a) $x$ = 0.1, (b) $x$ = 0.2, (c) $x$ = 0.3, (d) $x$ = 0.4. (Scale of 1 μm is for all images shown in (d)).

## 3.2. Permittivity Analysis

The temperature dependent permittivity (real) measured at 1 kHz for (100-$x$)LNN-$x$BT ($x$ = 10, 20, 30, 40) ceramics is shown in figure 2. The permittivity peaks associated with ferroelectric-paraelectric phase transition ($T_m$) are found at $T_m$ = 265, 125 °C for $x$ = 10, 20 respectively, whereas $T_m$ = -15, -90 °C for $x$ = 30, 40 respectively (shown in inset of figure 2) [1].

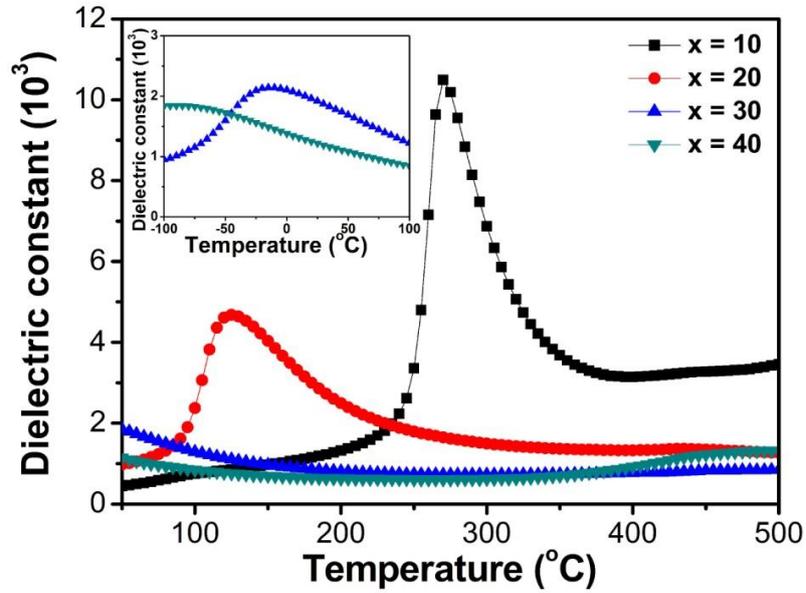

**Figure 2:** Temperature dependent dielectric constant at 1 kHz for the temperature range 50-500 °C for (100-$x$)LNN-$x$BT ceramics. Inset shows a temperature variation for the temperature range of -100 to 100 °C for $x$ = 30, 40.

Figure 3 depicts the frequency dependence (0.01 to 1 MHz) of real and imaginary part of dielectric permittivity ($\varepsilon'$ and $\varepsilon''$ respectively) on a log-log scale for the temperature range 50 to 500 °C for (100-$x$)LNN-$x$BT ($x$ = 10, 20, 30, 40) ceramics. While figure 4 shows the same graph that is plotted for the temperature range -100 to 100 °C, depending on the low transition temperatures of composition, $x$ = 30, 40. As seen in figure 3 and 4, both the parameters ($\varepsilon'$ and $\varepsilon''$) show strong low frequency dispersion in dielectric permittivity (LFDD), a typical characteristic of high-temperature behavior. According to Jonscher, some of the slowly moving hopping charge carriers (localized) may jump over several consecutive sites and not just the adjacent sites, leading to hopping conduction. However, when localized charges are transferred by applying field to the adjacent sites (short range hopping) it leads to dielectric relaxation. In (100-$x$)LNN-$x$BT ceramics, a LFDD is found both above and below the transition temperature ($T_m$) and no loss peak in $\varepsilon''(f)$ was observed. This is usually a carrier dominated dielectric response where charged carriers may hop in the system and their collective interaction lead to a LFDD [10]. At relatively higher frequencies (but ≤ 1 MHz), charged carriers can not follow the external field, therefore, permittivity mainly results from intrinsic polarization and become almost frequency independent as seen in figure 3 and 4. On the other hand, dipolar response (temperature dependent) that is present in the system for the entire frequency range shows up at the higher frequencies. This is evident from the observation that for each composition at higher frequencies as temperature approaches to $T_m$, $\varepsilon'(T)$ shows a maximum value indicated by an arrow for $x = 10, 20$ in figure 3 and for $x = 30, 40$ in figure 4.

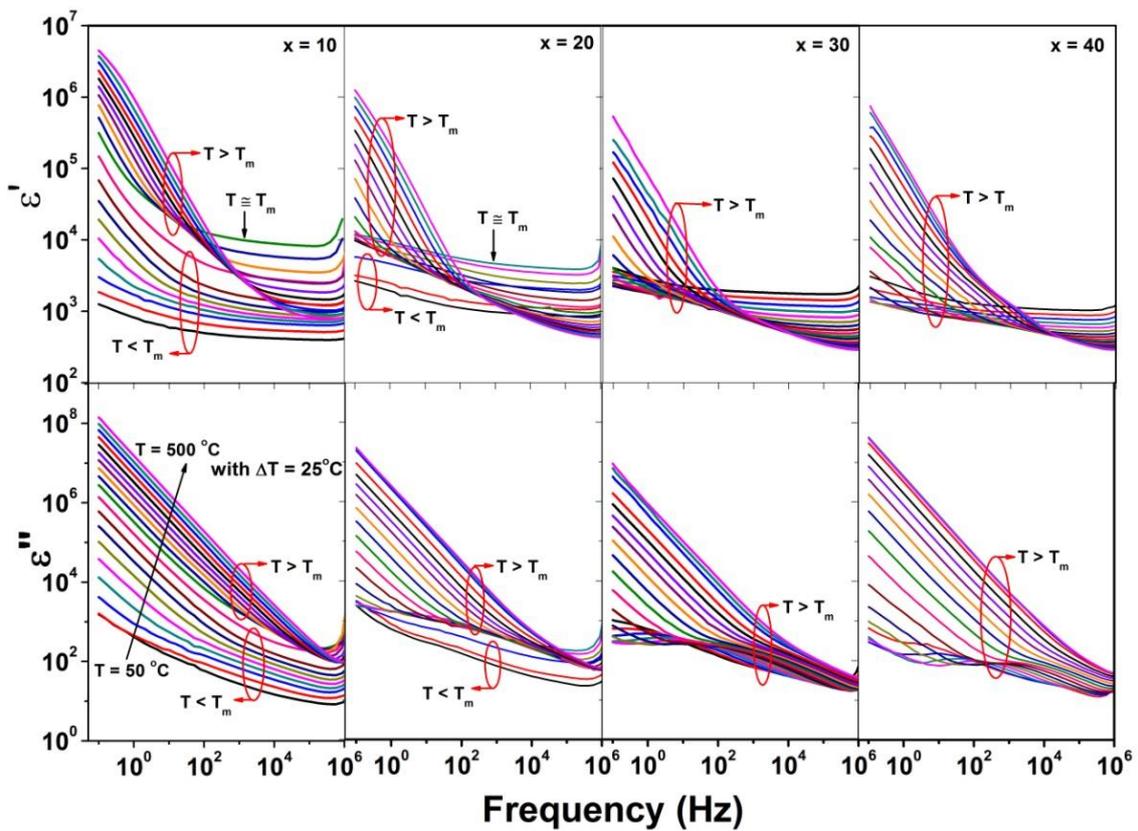

**Figure 3** Variation of real and imaginary part of dielectric constant with frequency for the temperature range 50-500 °C for $x = 10, 20, 30, 40$

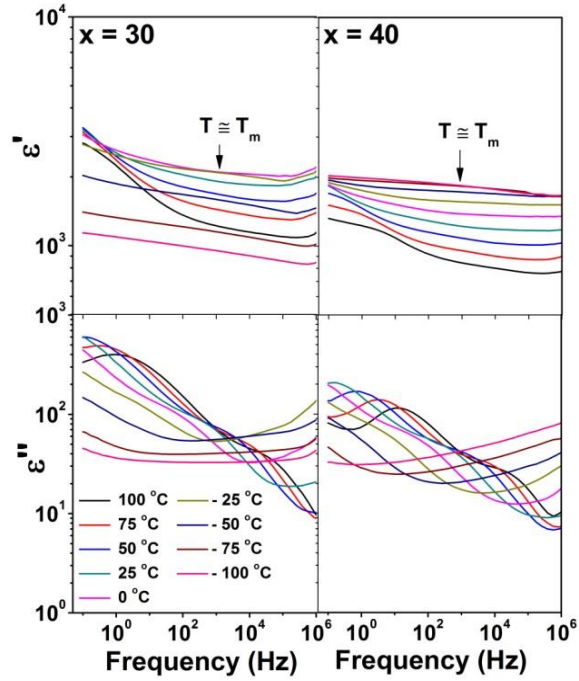

**Figure 4** Variation of real and imaginary part of dielectric constant with frequency for the temperature range -100 to 100 $^{o}$C for $x$ = 30, 40 with a step of 25 $^{o}$C for (100-$x$)LNN-$x$BT ceramics.

All these frequency dependent dielectric permittivity behavior can be well understood using Jonscher's fraction power law, the universal relaxation law ($\varepsilon'$, $\varepsilon'' \propto \omega^{n-1}$) [11]. To look into the details of relaxation behavior in these materials, the same data shown in figure 3 and 4 is replotted only at two representative temperatures (below and above $T_m$ each) as depicts in figure 5. Two different regions in $\varepsilon'(f)$ can be observed: as frequency increases from low frequency side, a linear decrease with slope close to -1 in the low frequency region and an almost frequency independent plateau region with a slope ($n$-1) in the higher frequencies. For $\varepsilon''(f)$, at higher frequencies a relatively shallow fractional power law with slope $-(1-n)$ [$0.1 \leq (1-n) \leq 0.3$] is observed. A linear increase with slope close to -1 is observed as the frequency approaches to lower side which could be due to the influence of conductivity. Jonscher identified another form of dielectric relaxation which shows considerable conductivity known as quasi-DC process (QDC). In QDC process, the universal power law follows [12]:

$$\varepsilon' \propto \varepsilon'' \propto \omega^{n_1 - 1} \text{ for } \omega \gg \omega_c, n_1 > 0.6 \quad (1)$$

$$\varepsilon' \propto \varepsilon'' \propto \omega^{n_2 - 1} \text{ for } \omega \ll \omega_c, n_2 < 0.3 \quad (2)$$

Where $\omega_c$ is the critical frequency at which both the curves $\varepsilon'$, $\varepsilon''$ vs $f$ intersect each other and below which both follow a power law. In (100-$x$)LNN-$x$BT, as seen in figure 5, it is very clear that the LFDD is associated with a QDC process where the dipolar orientation or charge carrier transition occurs necessarily by discrete movements and that every orientation or transition contributes to $\varepsilon'$ makes a proportional contribution to $\varepsilon''$ [12]. This is completely contradictory to the DC conduction where $\varepsilon'$ remain constant and $\varepsilon''$ rises steadily as the frequency approaches towards lower frequencies. The charge species in (100-$x$)LNN-$x$BT could be cationic vacancies $V'_{Li/Na}$, $Ti'_{Nb}$ and oxygen-defect-related complexes ($V^{\bullet\bullet}_O$, ($Ti_{Nb}{}' - V^{\bullet\bullet}_O)^{\bullet}$) arising due to acceptor type doping [1]. These defects or defect complexes are highly mobile and plays an important role in LFDD and in hopping conduction. The defect dipoles act as they are pinned at one end rather than completely free to reorient and therefore swiping of dipoles considered as hopping of charges. For all compositions, the position of $\omega_c$ is found to shift towards lower frequency side as temperature decreases.

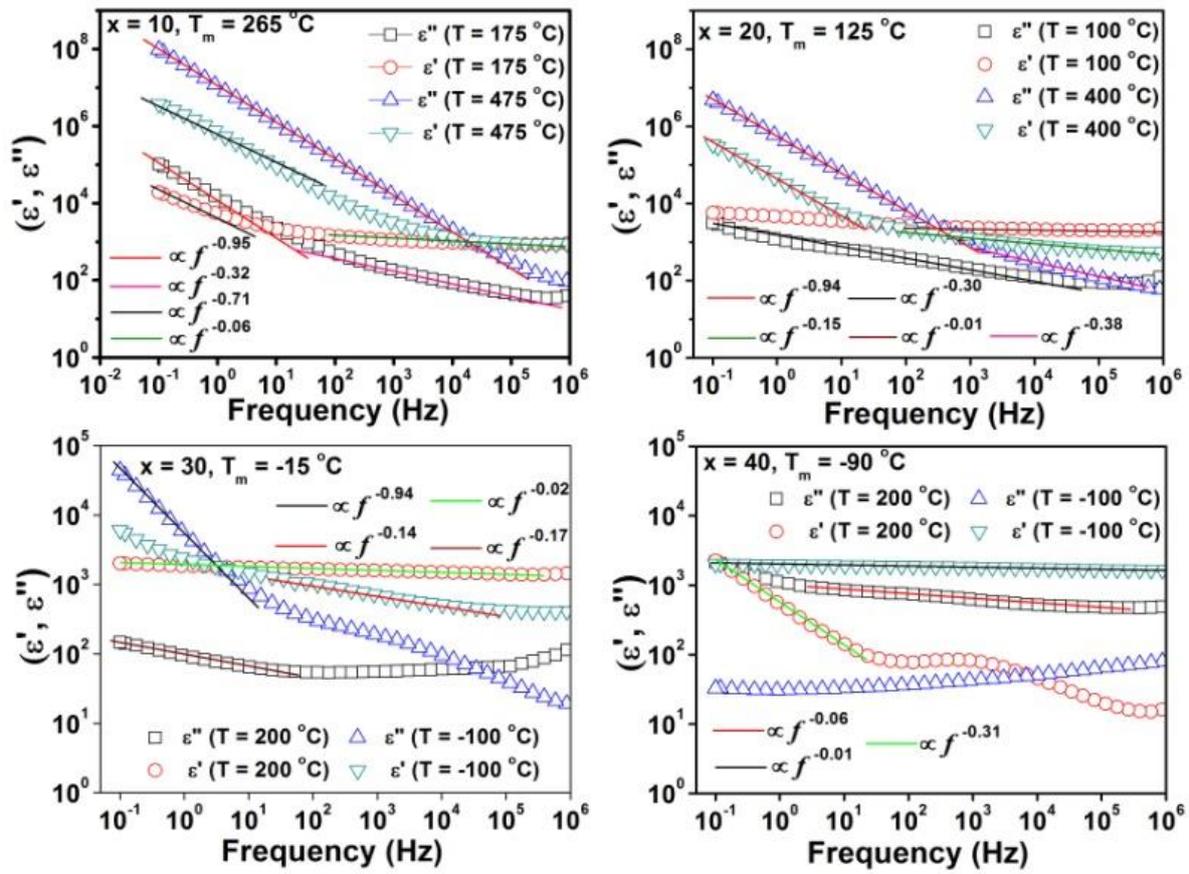

**Figure 5** Frequency dependence of real and imaginary part of dielectric constant at few representative temperatures below and above $T_m$. Solid lines represent the fitting using universal power law. A log-log plot provides a better look for the power frequency dependency of dielectric parameters.

### 3.3 Conductivity Analysis

The variation of AC conductivity ($\sigma_{ac}$) with frequency for the temperature range 50-500 $^oC$ of (100-$x$)LNN-$x$BT ceramics for the studied compositions ($x$ = 10, 20, 30, 40) are shown in figure 6. AC conductivity, $\sigma_{ac}$ has been found to obey Jonscher's power law:

$$\sigma_{ac}(\omega) = \sigma_{dc} + A(T)\omega^{n(T)} \tag{3}$$

where $\sigma_{dc}$ is the DC conductivity (frequency independent), $A(T)$ is a temperature-dependent constant, and $n(T)$ is also a temperature-dependent exponent lies in the range $0 < n < 1$ [13]. The term $A\omega^n$ characterizes the frequency dependency of conductivity and the dispersion phenomena. From figure 6, it is seen that for all the plots, $\sigma_{ac}$ increases with increase in frequency as characteristics of $\omega^n$, however, at higher temperatures and low frequencies $\sigma_{ac}$ show an almost frequency independent response ($\sigma_{dc}$) and show $\omega^n$ dependence only at higher frequency. Significant frequency dispersion in the conductivity at lower frequencies was observed for temperatures below ferro-paraelectric phase transition temperature ($T_m$). However, above $T_m$ and at higher frequencies, $\sigma_{ac}$ is found to merge for all temperatures, whereas below $T_m$ frequency dispersion still remains as marked by red circle in figure 6. This behavior suggests that conductivity mechanisms are different in ferroelectric and paraelectric phases and near the transition temperature ($T_m$).

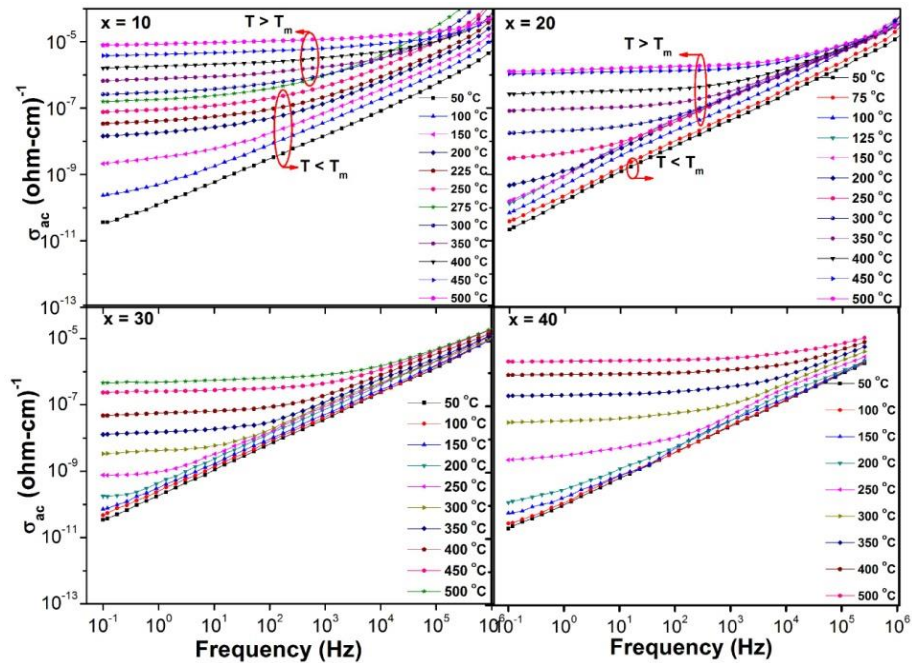

**Figure 6** Frequency variation of total conductivity at different temperatures for (100-$x$)LNN-$x$BT ($x$ = 5, 10, 20, 30) ceramics.

To understand the observed frequency dispersion and conductivity relaxation, Jonscher's universal power law (eqn. (3)) to the experimental data was fitted and from the fitting, $\sigma_{dc}$ values are estimated. However, attempts to fit eqn. (3) near $T_m$ was unsuccessful due to frequency dispersion in AC conductivity ($\sigma_{ac}$). Therefore, $\sigma_{ac}$ was fitted in the high frequency

(HF) and low frequency (LF) separately for the temperatures near $T_m$. Figure 7 shows fitting of eqn.(3) and the conductivity parameters for HF and LF regions for $x = 20$ (representative composition, $T_m = 125\ ^oC$) of $(100-x)$LNN-$x$BT ceramics. It is evident from figure 7(a) that for $\sigma_{ac}$ vs frequency recorded at $200^oC$ gives two values of each conductivity parameter, whereas for T = 500 $^oC$ conductivity parameters are same in both LF and HF. Figure 7(b) shows the variation of DC conductivity with 1000/T that obtained from the fitting of eqn. (3) at LF and HF. DC conductivity recorded at the lowest measured frequency (0.1 Hz) is also plotted. A significant difference in DC conductivity values ($\sigma_{DC}$ (HF), $\sigma_{DC}$ (LF), $\sigma_{DC}$ (0.1)) is observed at low-temperatures, while almost same values are seen in high temperatures. This confirms strong low frequency dispersion at low temperature and a dominant DC conductivity at higher temperatures.

In the low temperature ferroelectric phase (LT-FE) extrinsic conduction resulting from impurity ions present in the lattice dominates, whereas intrinsic ionic conduction results from the movement of the component ions in the high temperature paraelectric phase (HT-PE) [7, 8]. In high-temperature paraelectric (HT-PE) region the macroscopic polarization disappears, while new charges are generated due to thermal agitation. This leads to higher conductivity similar to ionic conductors. At sufficiently higher temperature most of the materials show increased movement of ions, either intrinsic to their lattice (ionic solids) or extrinsic due to impurities. This generally leads to DC conduction, however, in many cases there is evidence of low frequency dispersion in data which on superficial analysis considered as DC. Ionic conduction relies on the formation of lattice defects under the action of thermal excitation, thus creating vacancies through which ionic motion may proceed under the action of external electric field.

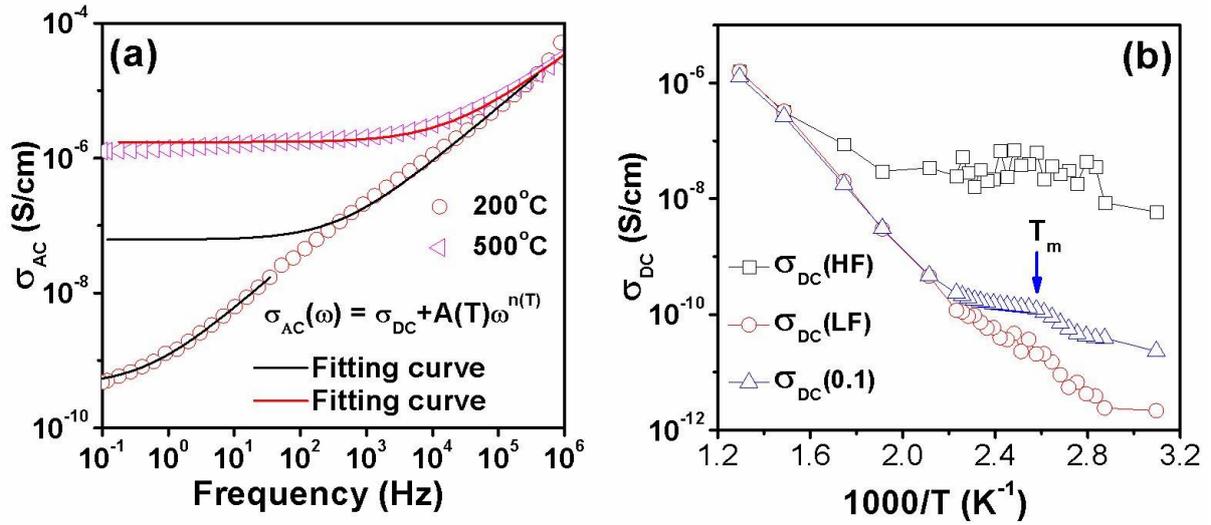

**Figure 7** Plot of (a) Jonscher's power law to determine $\sigma_{dc}$; solid line represents Jonscher's power law; (b) Variation of $\sigma_{dc}$ with *1000/T* (Arrhenius plot) for (100-*x*)LNN-*x*BT ceramics for *x* = 20 as a representative composition. More temperature data are shown near $T_m$ (= 125 °C).

DC conductivity ($\sigma_{dc}$) is a thermally activated process obey Arrhenius relation, $\sigma_{dc} = \sigma_0 \exp(-E_a/kT)$, where $\sigma_0$ is the pre-exponential factor and a characteristic of the materials, $k$ is Boltzmann's constant, $T$ is the absolute temperature and $E_a$ is the activation energy associated with the conduction mechanism. The activation energy of DC conductivity ($E_a$) in the analyzed temperature range are estimated from the slope of $\ln(\sigma_{dc})$ vs $1000/T$ plot [14] and listed in table 1. As expected, two different values of $E_a$ are observed for two regions: low-temperature ferroelectric (LT-FE) and high-temperature paraelectric (HT-PE). The range of activation energies of DC conduction obtained for (100-*x*)LNN-*x*BT ceramics are arising due to hopping of charged defects. This could be well interpreted as thermal motion of doubly ionized oxygen vacancies or to the formation of defect dipoles between the acceptor ion and charge compensating oxygen vacancies [15, 16]. The transportation of highly mobile oxygen vacancies in perovskite at high temperature give rise to an activation energy of ~1 eV [15, 17] whereas, the activation energies for transportation of *A*- and *B*-site cation need much higher activation energy [18]. The *A*-site vacancies in are inevitable as *A*-site cations (Li$^+$,

$Na^+$) are highly volatile during processing and $Li^+$ may occupy random sites, which resulted in oxygen vacancies to maintain charge neutrality [19, 20]. When $BaTiO_3$ is doped into $(Li,Na)NbO_3$ the defect reaction can be expressed as,

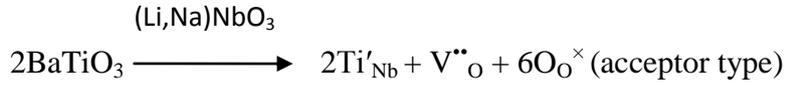

$$2BaTiO_3 \xrightarrow{(Li,Na)NbO_3} 2Ti'_{Nb} + V^{\bullet\bullet}_O + 6O_O^{\times} \text{ (acceptor type)}$$

Nobre et al. [21] calculated the activation energy of DC conductivity for pure LNN and found the values ranging 0.88-1.65 eV. The influence of structural changes across transition temperature ($T_m$) on the conductivity mechanism was explained in term of small polaron type for the pure LNN system. Similar conduction mechanism behavior is observed in the present study, however, the difference in calculated activation energy of DC conductivity could be attributed to the formation defect complex $(Ti_{Nb}' - V^{\bullet\bullet}_O)^{\bullet}$.

**Table 1** Variations of activation energy of DC conductivity of ferroelectric (FE) and paraelectric (PE) phase with BT content of (100-$x$)LNN-$x$BT ceramics.

| $x$ (mol%) | $E_a$ (PE) (eV) | $E_a$ (FE) (eV) |
|---|---|---|
| 10.0 | 0.64 | 0.74 |
| 20.0 | 0.91 | 1.11 |
| 30.0 | 0.83 | 0.98 |
| 40.0 | 1.20 | 1.28 |

Therefore, conductivity relaxation mechanisms were anticipated to occur inside the grain, conditioned by potential barriers due to space charge accumulated at the grain boundaries. Therefore impedance spectroscopy has been used to make a distinction between grain and grain boundary effect in the microscopic processes which involves dielectric relaxation and long range conductivity (i.e., localized and non-localized conduction respectively)

## 3.4 Impedance and Modulus Analysis

The impedance data (Z″ vs Z′) of (100-$x$)LNN-$x$BT ceramics for $x$ = 20 (results for other compositions were similar and omitted for briefness) in the Nyquist diagram for different selected temperatures are depicted in figure 8. A straight line has been observed with a larger slope at lower temperature indicating the insulating behavior of the samples. Above the transition-temperature ($T_m$), the slope of the curve decreases, bending down towards the real axis and become semicircle at higher temperature. This indicates the presence of both localized and non-localized conduction mechanism.

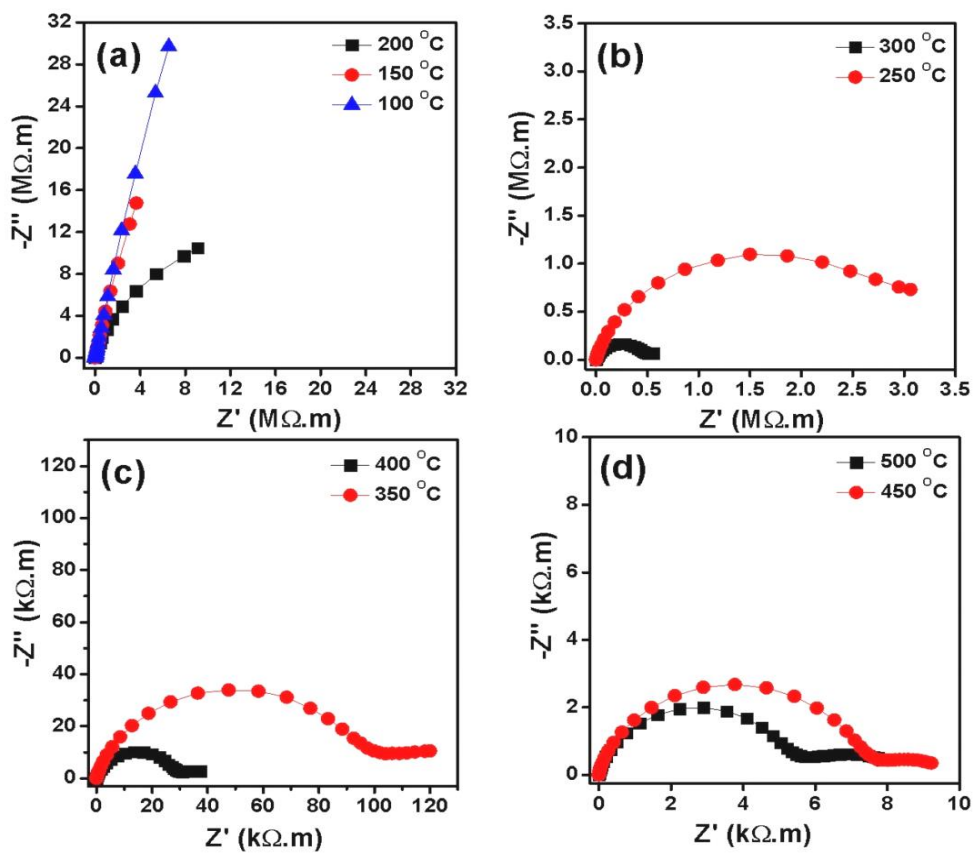

**Figure 8** Complex impedance plot (Z″ vs Z′) of (100-$x$)LNN-$x$BT ceramics for $x$ = 20 at temperatures range of 100- 500 °C. (Similar behavior was found for other compositions).

The complex impedance (Z$^*$) plot at 500 °C and 450 °C for $x$ = 30 as a representative composition of (100-$x$)LNN-$x$BT ceramics is shown in figure 9(a). The plots show typical

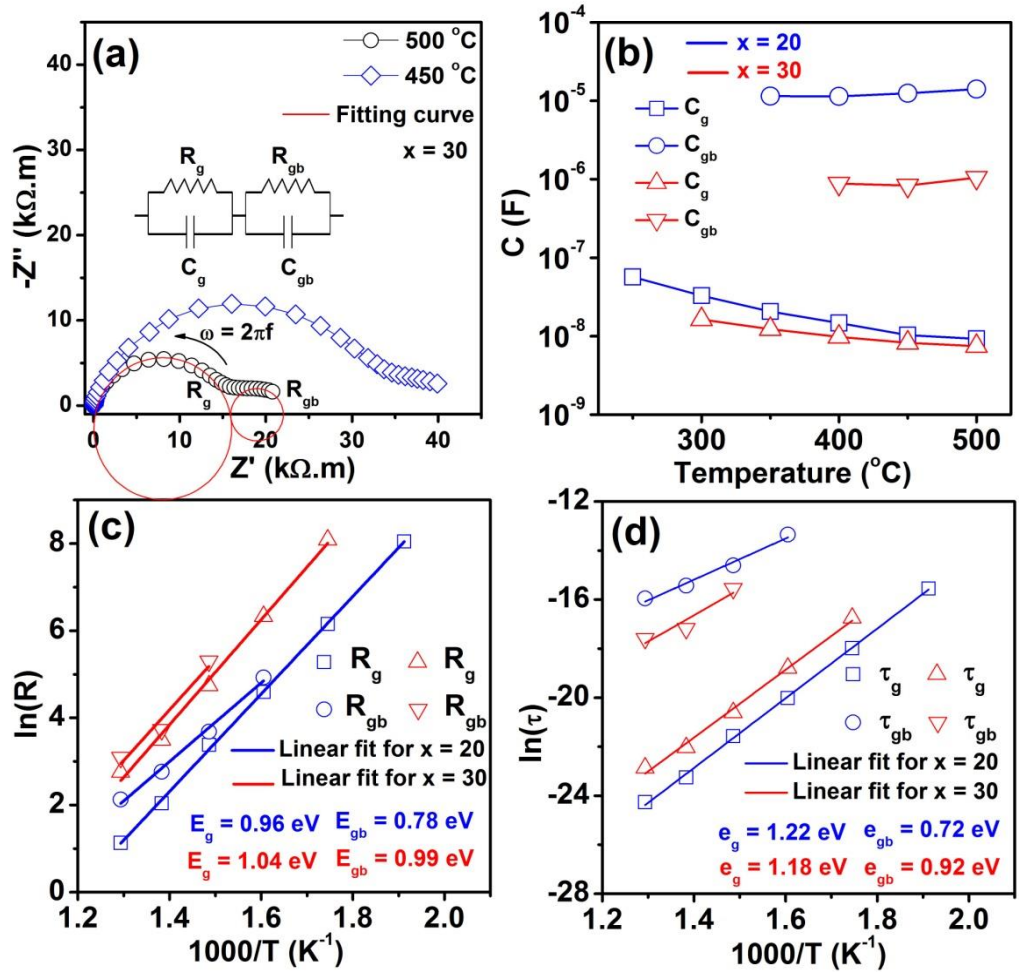

**Figure 9** The complex impedance plot (Z″ vs Z′) of (100-x)LNN-xBT ceramics for x = 30 at temperatures 500 and 450 °C. Two parallel RC equivalent circuits corresponding to grain and grain-boundary response is shown in inset. (A similar behavior was found for other compositions)

semicircle response, however slightly depressed instead of centered on the abscissa axis. The most accepted interpretation of the depressed semicircle is attributed to statistical distribution of relaxation times [22, 23]. Debye model suggests a single relaxation time gives rise to an ideal semicircle centered on the abscissa axis, however, in the present diagram shapes are due to at least two phenomena of relaxation with different relaxation frequencies which obeys a non-Debye model. This suggests that both the contributions follow Cole-Cole model. Thereby, based on previous studies [23, 24] and observations of present data, the semicircles as seen in complex impedance ($Z^*$) plot has been assigned to two semicircles, one in the low-frequency regime that is related to the grain boundary effect, while the high-frequency semicircle corresponds to the (grain) bulk relaxation. These observations are in good agreement with the SEM micrographs which clearly depicts a distinct grain and grain boundary (Fig. 1). The equivalent circuit shown in the inset of figure 9(a) is frequently used to describe the series of interconnected grain and grain-boundary phenomena [22]. In this circuit model, the electrical response from grain and grain-boundary are assumed to follow Debye-like behavior represented by two RC circuits, which in general, do not represent the real situation.

To determine the resistance-capacitance response of grains ($R_g$, $C_g$) and grain-boundaries ($R_{gb}$, $C_{gb}$), the impedance data were analyzed by means of the complex nonlinear least-square method which fits the equivalent circuits, consisting two parallel R-C circuits in series, to the Nyquist plot. The semicircles can be well resolved into two semicircles and fitted with two semi-circular curves. An illustration of experimental data and generated fitting curve can be seen in figure 9(a) and the values of $R_g$, $R_{gb}$, $C_g$, $C_{gb}$ for compositions $x$ = 20, 30 for few selected temperatures (500, 450, 400, 350, 250 $^o$C) are listed in table 2 for analysis. At lower temperatures the semicircles are not well resolved for $R_g$, $R_{gb}$. It may be noted from table 2 that, the grain boundary resistance ($R_{gb}$) is higher than the bulk or grain resistance ($R_g$) for all composition (*x*). It shows that grain boundaries are more resistive than grains and the obtained range of values for $R_g$, $R_{gb}$ confirm the presence of electrically inhomogeneous microstructure consisting of insulating grain boundaries and semiconducting grains. Thus, there is a chance of the formation of a barrier layer capacitor between semiconducting grains and insulating grain boundaries which produce a large polarization across grain boundaries leading to increased value of ε′ at low frequencies as found in figure 4 [25-27]. Both $R_g$ and $R_{gb}$ also found to increase with a decrease in temperature in PE-phase suggests an increase in conductivity with temperature.

For all the compositions, grain boundary ($C_{gb}$) capacitance is found higher than grain or bulk ($C_g$) capacitance. The values of $C_g$ increases with decrease in temperature while, $C_{gb}$ decreases with decrease in temperature in PE-phase. This result is in agreement with temperature-dependent dielectric permittivity (capacitance) measurement where above ferroelectric-paraelectric transition temperature dielectric permittivity decreases with rise in temperature [28]. This suggests that there may not be any interfacial effect [29]. As the conductivities for grains and grain boundaries are different owing to their respective underlying mechanisms, thereby they relax also at different frequencies. The relaxation frequency or peak frequency ($\omega_{max} = 2\pi f_{max}$) of the semicircle can be expressed as: $\omega_{max}\tau = 1 = RC$, where $\tau$ is the relaxation time, R and C represent the resistance and capacitance of grain and grain boundaries. The relaxation time ($\tau$) for grain and grain boundaries ($\tau_g$, $\tau_{gb}$) are calculated and listed in table 2. The temperature dependence of the relaxation time ($\tau$) and resistance (R) follows Arrhenius law,

$\tau = \tau_o \exp(-e_a/k_B T)$

$R = R_o \exp(-E_a/k_B T)$

where $R_o$, $\tau_o$ are the pre-exponential factors, $k_B$ is Boltzmann constant, T is the absolute temperature and $E_a$, $e_a$ are the activation energies for grain and grain boundary conduction and relaxation respectively. Activation energies for grain and grain boundary conduction ($E_g$, $E_{gb}$) and relaxation ($e_g$, $e_{gb}$) are calculated from the ln($\tau$) and ln(R) vs 1000/T plot respectively as shown in figure 9(a) and 9(b) for $x$ = 30, 40. The activation energies for relaxation are generally associated with a free energy for migration of charge carriers and hopping of these between the adjacent lattice site whereas, for conduction it require free energy for both creation and migration of these charges over a long distance. From figure 9, it is clear that $E_g$, $E_{gb}$ > $e_g$, $e_{gb}$ which signify that a long range movement of defects require high thermal energy than short range hopping. The ranges of activation energy are associated with the conduction mechanism dominated by grain boundary conduction through hopping electron created through oxygen vacancies.

**Table 2** Equivalent circuit component values of $R_g$, $R_{gb}$, $C_g$, $C_{gb}$, $\tau_g$, $\tau_{gb}$ at different temperatures for (100-$x$)LNN-$x$BT ceramics for $x$ = 20 and 30.

|  |  Temperature (°C) | 500 | 450 | 400 | 350 | 300 | 250 |
|---|---|---|---|---|---|---|---|
| $x$ = 20 | $R_g$ (kΩ) | 3.13 | 7.69 | 29.34 | 98.69 | 470.20 | 3120.30 |
|  | $R_{gb}$ (kΩ) | 8.39 | 15.84 | 39.86 | 138.20 | - | - |
|  | $C_g$ (nF) | 9.24 | 10.30 | 14.70 | 20.60 | 32.90 | 56.70 |
|  | $C_{gb}$ (μF) | 14.06 | 12.49 | 11.41 | 11.52 | - | - |
|  | $\tau_g$ (ns) | 0.0289 | 0.0792 | 0.431 | 2.033 | 15.469 | 176.921 |
|  | $\tau_{gb}$ (μs) | 0.1179 | 0.1978 | 0.4548 | 1.592 | - | - |
| $x$ = 30 | $R_g$ (kΩ) | 15.7 | 32.8 | 114.6 | 561.8 | 3247.2 | - |
|  | $R_{gb}$ (kΩ) | 22.0 | 41.6 | 201.0 | - | - | - |
|  | $C_g$ (nF) | 7.49 | 8.22 | 9.8 | 12.3 | 16.3 | - |
|  | $C_{gb}$ (μF) | 1.05 | 0.83 | 0.88 | - | - | - |
|  | $\tau_g$ (ns) | 0.1176 | 0.269 | 1.123 | 6.910 | 52.929 | - |
|  | $\tau_{gb}$ (μs) | 0.0231 | 0.0345 | 0.176 | - | - | - |

Figure 10 depicts the combined plot of Z″ and M″ as a function of measured frequency for (100-$x$)LNN-$x$BT ceramics at temperature above their respective ferro-paraelectric transition temperature ($T_m$) [28]. This is particularly useful to understand electrical behavior since Z″ plots highlight phenomena characterized by largest resistance, and M″ plots identify electrical responses with smallest capacitance [30, 31]. These plots also help in distinguishing the difference between localized (i.e. defect relaxation) and non-localized conduction (i.e. ionic and electronic conductivity) processes within the bulk of the material as both the processes give rise to the same geometric capacitance [9, 32, 33]. All the plots show that Z″ value increases initially, attain a peak ($Z_{max}″$) and then decreased as frequency increases at all measured temperatures. However, at high-temperature a small peak appears at low frequency region for all the compositions which is due to the grain boundary contribution. At high-frequency, all Z″ curves merge due to release of space charges, as at high-frequency a less time is available for space charges to relax back, hence reducing space charge polarization leading to merging of all curves. The broad and asymmetric shapes of the peaks are indicative of a spread in relaxation times, thereby consisting of multiple relaxations that are consistent with diffuse character of (100-$x$)LNN-$x$BT ceramics. The asymmetric broadening especially at higher temperature reveals the presence of electrical phenomena with a spread of relaxation times that arises due to cationic disorder at both *A*- and *B*-sites [34]. This also reveals that there is a deviation from the ideal Debye behavior and presence of temperature-dependent electrical relaxation phenomena. The relaxation process arises due to the presence of immobile species at low temperature and defects at higher temperature [35]. The relaxation peaks observed in Z″ is either related to short range hopping (localized) (i.e., defect relaxation) or long range (non-localized) (i.e., ionic and electronic conductivity) conduction mechanism [33]. The values maximum of Z″ (magnitude of $Z_{max}″$) is found to decreases with rise in temperature which is due to decrease in resistance of the materials. It indicates an increase in the resistive property, as $Z_{max}'' = R/2$. The position of the peak ($Z_{max}''$) shifts systematically towards the higher frequency side with the rise in temperature indicating a decrease in relaxation time.

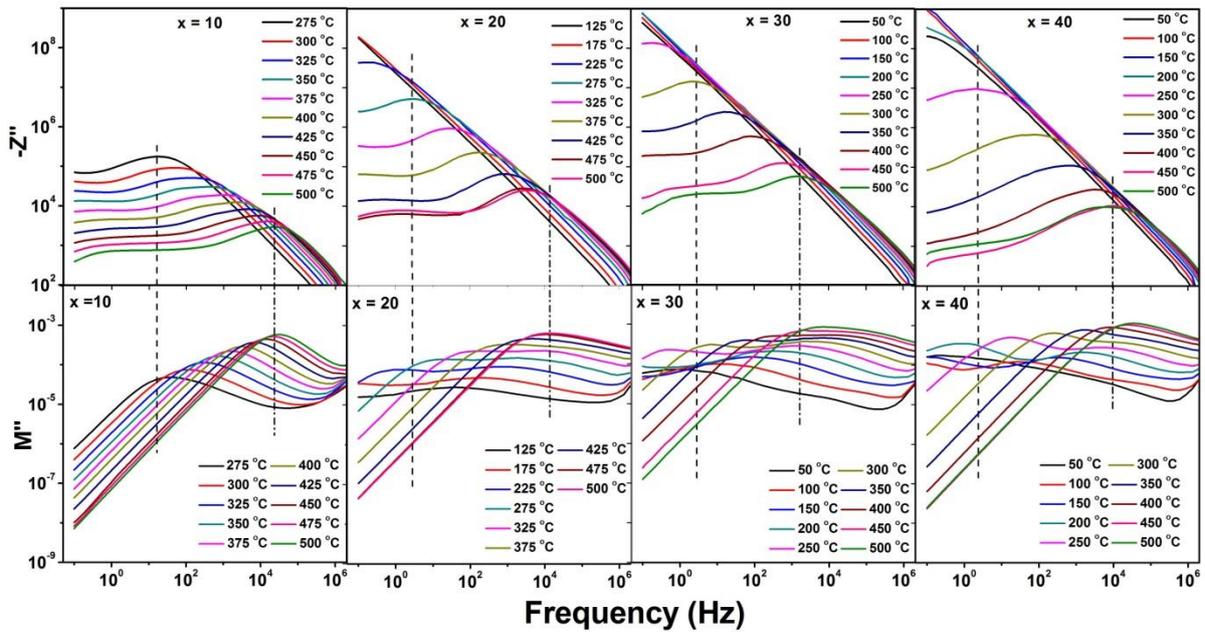

**Figure 10** Variation of imaginary part of impedance (Z″) and electric modulus (M″) with frequency at different temperatures for (100-$x$)LNN-$x$BT ($x$ = 10, 20, 30, 40) ceramics.

In electric modulus (M″) plot, the peak positions in M″ shift towards higher frequency as temperature increases, which is pointing to a temperature dependent electrical relaxation phenomenon in the material. The value of maximum of M″ ($M_{max}'' = \varepsilon_o/2C$) also increases with increase in temperature indicating a decrease in capacitance with temperature. This peak is associated with bulk or grain interior effect of the materials. A small shoulder is observed in the frequency range between 10-10$^5$ Hz at lower temperatures for all compositions except $x$ = 10. The position of the shoulder shifts towards higher frequency side with the rise in temperature and moves above the measurable frequency range. The shoulder is believed to arise in (100-$x$)LNN-$x$BT ceramics due to the relaxation of highly polarizable small entity (PNRs).

The relaxation process due to short range hopping, peaks of Z″ and M″ occur at different frequencies while for long range they occur at the same frequency. In (100-$x$)LNN-$x$BT ceramics, it is observed that there is a slight separation between the position of Z″ and M″ peaks (figure 10), which suggests a short range hopping (i. e., localized) conduction mechanism [36].

**4. Conclusions (incomplete)**

Dielectric behavior and conductivity mechanism in (100-$x$)LNN-*XBT* ceramics were investigated from the temperature and frequency dependent permittivity data. Low frequency dispersion in dielectric permittivity (LFDD), a typical characteristic of high-temperature behavior was observed both below and above the ferro-paraelectric phase transition temperature, $T_m$. In relation to the occurrence of LFDD in (100-$x$) LNN-$XBT$ ceramics, cation vacancies ($V'_{Li/Na}$, $Ti'_{Nb}$ and oxygen-defect-related complexes ($V^{\bullet\bullet}_O$, $(Ti_{Nb}' - V^{\bullet\bullet}_O)^{\bullet}$) were found to play an important role in LFDD and hopping conduction. The collective interaction of these defects leads to LFDD and high frequency response is dominated by bipolar response. The dielectric relaxation follows Jonscher universal law, however LFDD is found to be associated with QCD process. The frequency dispersion and conductivity relaxation near $T_m$ is also investigated and two separate mechanisms of conduction were founded: LFDD at low temperature and DC conduction at high temperature found. The activation energies of DC conduction confirm the mechanism as the thermal motion of doubly ionized oxygen vacancies or to the formation of defect dipoles between the acceptor ion and charge compensating oxygen vacancies.

**Acknowledgement**